\documentclass[%
 reprint,onecolumn,
 amsmath,amssymb,
 aip, 12pt
]{revtex4-2}
\usepackage{xcolor}
\usepackage{graphicx}
\usepackage{dcolumn}
\usepackage{bm}
\usepackage{hyperref}
\usepackage{setspace}
\usepackage{threeparttablex}
\usepackage{booktabs}
\usepackage{array}
\usepackage[title]{appendix}
\newcommand{\bra}[1]{\ensuremath{\left\langle#1\right|}}
\newcommand{\ket}[1]{\ensuremath{\left|#1\right\rangle}}
\newcommand{\ketbra}[2]{\ensuremath{\left|#1\right\rangle\left\langle#2\right|}}

\newcolumntype{P}[1]{>{\centering\arraybackslash}p{#1}}
\begin{document}

\title{On the nature of two-photon transitions for a Collection of Molecules in a Fabry-Perot Cavity }
\author{Zeyu Zhou}
\affiliation{Department of Chemistry, University of Pennsylvania, 231 South 34th Street, Philadelphia, Pennsylvania 19104, United States}
\author{Hsing-Ta Chen}
\affiliation{Department of Chemistry, University of Pennsylvania, 231 South 34th Street, Philadelphia, Pennsylvania 19104, United States}
\affiliation{Department of Chemistry and Biochemistry, 251 Nieuwland Science Hall, Notre Dame, Indiana 46556, United States}
\author{Maxim Sukharev}
\affiliation{Department of Physics, Arizona State University, Tempe, Arizona 85287, United States}
\affiliation{College of Integrative Sciences and Arts, Arizona State University, Mesa, Arizona 85212, United States}
\author{Joseph E. Subotnik}
\affiliation{Department of Chemistry, University of Pennsylvania, 231 South 34th Street, Philadelphia, Pennsylvania 19104, United States}
\author{Abraham Nitzan}
\affiliation{Department of Chemistry, University of Pennsylvania, 231 South 34th Street, Philadelphia, Pennsylvania 19104, United States}
\affiliation{Department of Physical Chemistry, School of Chemistry, The Raymond and Beverly Sackler Faculty of Exact Sciences and The Sackler Center for computational Molecular and Materials Science, Tel Aviv University, Tel Aviv 6997801, Israel}

\date{\today}

\begin{abstract}
We investigate the effect of a cavity on  nonlinear two-photon transitions of a molecular system and how such an effect depends on the cavity quality factor, the field enhancement and the possibility of dephasing. We find that the molecular response to strong light fields in a cavity with variable quality factor can be understood as arising from a balance between (i) the ability of the cavity to enhance the field of an external probe and promote multiphoton transitions more easily and (ii) the fact that the strict selection rules on multiphoton transitions in a cavity support only one resonant frequency within the excitation range.
Although our simulations use a classical-level description of the radiation field (i.e. we solve Maxwell-Bloch or Maxwell-Liouville equations within the Ehrenfest approximation for the field-molecule interaction), based on experience with this level of approximation in past studies of plasmonic and polaritonic systems, we believe that our results are valid over a wide range of external probing. 
\end{abstract}
\maketitle

\newpage

\section{Introduction}
Microcavities, including  Fabry-Perot and plasmonic cavities, are well known for changing the local density of radiative states within the cavity and modifying the couplings between vibrational/electronic states to light, leading to the Purcell effect in the weak coupling regime and the Rabi splitting in the strong coupling regime at room temperature.\cite{bayer2001coupling,  walther2006cavity,hutchison2012modifying,  ebbesen2016hybrid, herrera2016cavity,kavokin2017microcavities, ribeiro2018polariton, frisk2019ultrastrong, hertzog2019strong} These alterations of the radiative couplings can result in drastic changes to the energies of light-matter states and energy flow pathways;  the responses of molecules inside a cavity towards external light sources can be very different from the response of bare molecules in solution.\cite{gerard1998enhanced, moreau2001single, unitt2005polarization, vuvckovic2003enhanced, muller2006self, ren2012spin, feist2015extraordinary, schachenmayer2015cavity, hagenmuller2017cavity, garcia2017long, saez2018organic, groenhof2018coherent, rozenman2018long, yuen2018molecular, galego2019cavity, herrera2020molecular, li2022molecular}

Even though numerically exact simulations of quantum dynamics are exponentially difficult, simulations of a classical electromagnetic field (using finite-difference time-domain (FDTD) method) combined with a quantum, mean-field description of the matter subsystem (usually comprising a set of two-level emitters, representing quantum dots or Rydberg atoms) have shown that a host of interesting cavity effects can in fact be captured by mixed quantum-classical algorithms.\cite{taflove2005computational, teixeira2007fdtd, mcmahon2007tailoring, zhao2008methods, puthumpally2014dipole, sukharev2017optics, you2019nonlinear, sidler2020polaritonic, tancogne2020octopus, sukharev2021second}
As far as one photon processes are concerned,  the Purcell effect arises semiclassically from constructive/destructive interference during multiple reflections of the electromagnetic (EM) field by the cavity mirrors; the Rabi splitting arises from the continuous oscillation of energy between the EM field and the molecules inside the cavity; cavity leakage/finite polariton lifetime can also be observed semiclassically if we simulate finite-size cavity mirrors with a complex-valued dielectric constant.
Nonlinear effects can also be observed in semiclassical simulations, for example, the Rabi splitting contraction results from ground state depletion can be simulated. 
\cite{li2022polariton, son2022energy}

Now, despite the power of FDTD approach, within the realm of polaritonics and cavity effects, most research today relies on model Hamiltonians and/or simulations with a single cavity mode.  Such approaches can often capture a great deal of physics, but one sacrifices the ability to capture quite a few EM observables.
For instance, an important example for this paper is the recent CavityMD study \cite{li2021cavity, li2022energy} demonstrating that, under illumination by a strong EM field, if a lower vibrational polariton state is tuned to half the frequency difference a $0\rightarrow 2$ vibrational excitation, exciting this  polariton state can enhance a nonlinear transition  by facilitating the energy transfer that arises from the intermolecular couplings between molecules inside the cavity.  This hypothetical numerical experiment suggests that solvent polariton excitations in a cavity can efficiently pump solutes and potentially promote chemical reactions.  The results of ref. \citenum{li2022energy}, however, were limited to absorption and could not easily address transmission, which is the most easily experimentally observed measurement. 
Here we present a mixed quantum-classical FDTD calculation that directly addresses this observable. \cite{soljavcic2004enhancement, delor2014toward} We note that, in order to work with the largest signal possible, we will simulate electronic transitions (instead of vibrational transitions). 

To carry out this program, we consider a Fabry-Perot cavity formed by two gold mirrors with different mirror thicknesses (and hence different quality factors). Inside the cavity, there is a layer of 3-level molecules (which are considered the solute) and the two-photon nonlinear transitions between the ground and the second excited states are calibrated to be resonant with the incoming laser field. 
The solvent is represented by two layers of 2-level molecules. Their interaction with the cavity mode leads to the formation of the lower/upper polariton states, thereby modifying the energy at which the EM field can be transmitted efficiently through the cavity. 
Next, we pump the system with external continuous wave (CW) light and compare the simulation results inside and outside of the gold cavity. 
Our simulations show $(i)$ that within a Maxwell-Bloch mean field treatment of the light-matter interaction, there is a field enhancement in the classical cavity and this field enhancement is the primary reason for altering two-photon nonlinear transitions. 
Furthermore, $(ii)$ the cavity is able to suppress non-resonant transitions between the second molecular excited state and the first excited state (as well as between the first excited state and the ground state)  as a manifestation of the Purcell effect. 
Hence, the resonant 2-photon nonlinear transitions are effectively shielded by the cavity mirrors, resulting in a prolonged Rabi oscillation, in contrast to the scenarios outside the cavity.  
$(iii)$ Increasing the cavity quality factor by making the gold mirrors thicker, we observe a turnover in the rate of the 2-photon nonlinear transitions. 
Finally, $(iv)$, we find that under constant illumination and in the presence of pure dephasing, the Rabi oscillations in the cavity are damped and the system reaches a time-independent steady state (which is often observed\cite{iorsh2012generic, schwartz2013polariton, dunkelberger2022vibration}).

This manuscript is arranged as follows. In section \ref{sec: two}, we present four model scenarios with layers of 2-level/3-level molecules and a cavity formed by two parallel gold mirrors.  In section \ref{sec: three}, we investigate population dynamics for these scenarios with cavities displaying different mirror thickness (quality factors) and we study how dephasing influences the two-photon transitions. In section \ref{sec: four}, we conclude and discuss the remaining questions.

\section{Model and Methods \label{sec: two}}
\subsection{Structure of the systems}
\begin{figure}[!ht]
    \includegraphics[width=14cm]{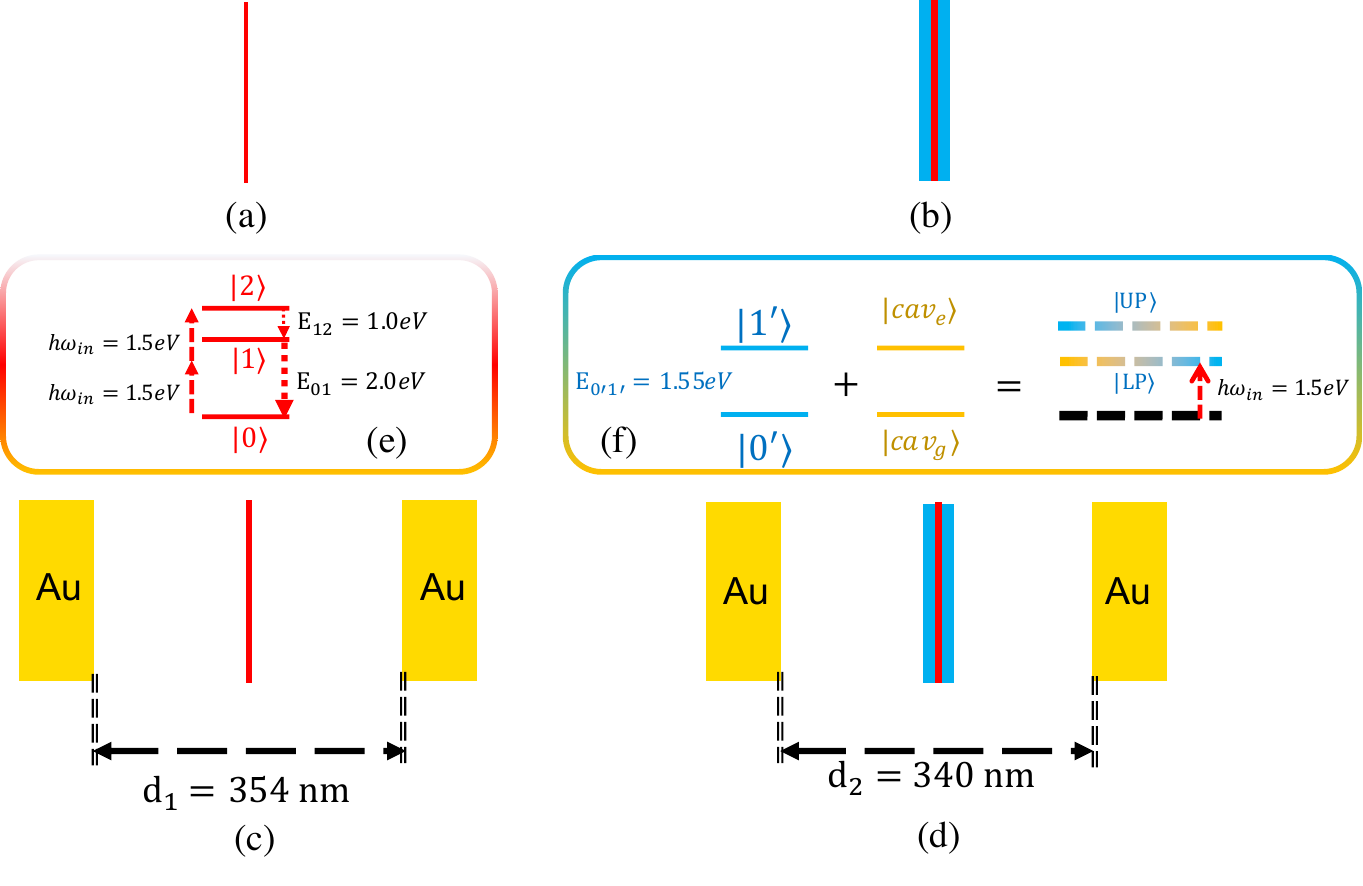}
    \caption{Sketches of the four scenarios that are studied in this work. (a) A single layer of 3-level molecules (a solute layer with states $\ket{0}, \ket{1}, \ket{2}$); (b) a single layer of 3-level molecules sandwiched by two layers of 2-level molecules (two solvent layers with states $\ket{0'}, \ket{1'}$); (c) Same as (a) with two surrounding planar gold mirrors forming a Fabry-Perot cavity; (d) Same as (b) but with two surrounding gold mirrors. All four scenarios are pumped by monochromatic external laser with angular frequency $\omega_{in} =1.5eV/\hbar$. In inset (e), we show the relevant transitions of the solute molecular layer alone. An external monochromatic laser matches the energy needed for 2-photon transition of the solute molecular layer. For scenario (c), the distance between the gold mirrors is $354$ nm, which is resonant  with the incoming driving frequency ($\hbar\omega_{in} = E_{cav} = 1.5$ eV). In inset (f), we draw the relevant lower and upper polariton states formed by strong interaction between cavity EM field and solvent molecular layers. For scenario (d), the cavity mode (with mirror distance $d_{2}=340$ nm) is on resonance with a bare solvent excitation ($E_{cav} = E_{0'1'} = 1.55$ eV), while the frequency of the lower polariton  matches the external monochromatic light ($E_{LP} = \hbar\omega_{in} = 1.50$ eV). }
    \label{fig: systemsketch}
\end{figure}
In this work, we consider four scenarios with continuous pumping by a laser source of frequency $\omega_{in}=E_{in}/\hbar$.
First, as shown in Fig \ref{fig: systemsketch}(a), we consider a single layer of 3-level (states $\ket{0}, \ket{1}$ and $\ket{2}$) solute molecules, with energy difference $E_{02} = 2E_{in}$ between levels $\ket{0}$ and $\ket{2}$. Level $\ket{1}$ is placed so that the separations between levels $\ket{0}$, $\ket{2}$ and level $\ket{1}$ do not overlap with the incoming frequency (in the simulations we take $E_{01} = 2.0eV \neq \hbar \omega_{in}$, $E_{12} = 1.0eV \neq\hbar\omega_{in}$) as shown in inset (e). Thus, at early times following excitation of the system, the population of $\ket{1}$ is expected to be small.
Second, as shown in Fig. \ref{fig: systemsketch} (b), we sandwich the 3-level molecules with two layers of 2-level ($\ket{0'}$, $\ket{1'}$) molecules (blue, 4.5 nm thickness each).  Here, the 3-level molecules are considered solute molecules and the 2-level molecules are considered solvent molecules.  (We assume that the solvent and solute molecules are kept together by a non-metallic structure that does not interact with the EM field.)

Next, as shown in Figs. \ref{fig: systemsketch} (c) and (d), we add gold mirrors (golden yellow, 50 nm thickness each) that encapsulate a microcavity. We use Drude model for the bulk gold to characterize its linear response to the EM field. In scenario (c), the distance between the two mirrors ($d_1$) is chosen such that the cavity is on resonance with the incoming driving frequency ($E_{cav} = \hbar\omega_{in}$). 
In scenario (d), the distance between the two mirrors ($d_2$) is taken such that the cavity is on resonance with solvent molecules, i.e. $E_{cav} = E_{0'1'}$. The solvent-cavity interaction now leads to the formation of upper/lower polariton states and we adjust $E_{0'1'}$ and $d_{2}$ (which controls $E_{cav}$) together to ensure $E_{LP} = \hbar\omega_{in}$; maintaining this condition clearly depends on the transition dipole moment for solvent molecules between $\ket{0'}$ and $\ket{1'}$) (i.e. the light-matter coupling).

\subsection{Maxwell-Bloch equations\label{subsec: maxwellequations}}
The interaction between the external EM pumping field is modeled in one-dimension. Thus, the one-dimensional Maxwell equations are
\begin{align}
    \frac{\partial B_{y}}{\partial t} &= -\frac{\partial E_{x}}{\partial{z}}\label{eq: mag}\\
    \epsilon_{0}\frac{\partial E_{x}}{\partial t} &= -\frac{\partial B_{y}}{\mu_{0}\partial{z}} - J_{x}\label{eq: elec}
\end{align}
Here, $z$ is the longitudinal coordinate that is perpendicular to the mirrors/slabs of nanolayers. $J_{x}$ is the polarization current
\begin{align}
    J_{x} = \frac{dP_{x}}{dt} = n_{0}\frac{d(\text{Tr}(\hat{\rho}\hat{\mu}_{x}))}{dt}
    \label{eq: current}
\end{align}
Here, $n_{0}$ is the number density, $\mu_{x}$ is a matrix of transition dipole moments between electronic states along $x$ direction. We simulate the EM equations above (eqs \ref{eq: mag} and \ref{eq: elec}) with an FDTD approach.

While the field is considered in the framework of the classical Maxwell equations, the molecular layers are described by their density matrices $\hat{\rho}$; $\hat{\rho}$ is $3\times 3$ for solute and $2\times 2$ for solvent molecules under a mean-field approximation. In other words, there is one density matrix for each grid point for each of the two molecular layers. The Hamiltonian (and the corresponding Liouvillian) are time-dependent because of the presence of the local time-dependent electric field.
\begin{align}
    i\hbar\frac{d}{dt}\hat{\rho} &= [\hat{H}(t), \hat{\rho}]
\end{align}
This field enters the Hamiltonian as the coupling between adjacent electronic states, i.e. $\bra{0}\hat{H}\ket{1} = \mu_{x}^{01}E_{x}(t)$.
In summary,
\begin{align}
\hat{H} =& E_{1}\ketbra{1}{1} + E_{2}\ketbra{2}{2} + \mu^{01}_{x} E_{x}(\ketbra{0}{1} + \ketbra{1}{0}) + \mu^{12}_{x} E_{x}(\ketbra{1}{2} + \ketbra{2}{1})
\nonumber
\\
\hat{H}' =& E_{1'}\ketbra{1'}{1'} + \mu^{0'1'}_{x} E_{x}(\ketbra{0'}{1'} + \ketbra{1'}{0'})
\end{align}
\subsection{Initial condition \label{subsec: initial condition}}
All molecular layers are initialized on the ground state ($\ket{0}$ and $\ket{0'}$, respectively). Near the edge of the simulation cell (e.g., far outside of any cavity), we drive one grid point as a continuous wave source with frequency $\omega_{in} = E_{in}/\hbar$ (which are chosen $1.5$ eV for all calculations below). 

\begin{align}
    E_{x}(z=z_{0}, t) = E_{0}\sin\omega_{in} t
\end{align}
Here, $z_{0}$ is near the edge of the simulation grid.
The EM field travels in both directions. Along one direction, the EM field reaches the material nanolayers;  along the other direction, the EM field is absorbed by the perfectly matched layer boundary conditions (PML) so that the molecular layers are not affected by unpredictable reflections at the boundary of the simulation system. \cite{taflove2005computational}

\section{Results \label{sec: three}}
In this section, we present numerical results from three perspectives. First, we study the role of the cavity in enhancing the EM field inside the cavity and in shielding the solute molecular nonlinear transitions. 
Second, we investigate how the 2-photon nonlinear transitions between ground state $\ket{0}$ and the second excited state $\ket{2}$ of the solute molecules depend on the cavity quality factor (which is controlled by the mirror thickness). Finally, we discuss how phenomenological pure dephasing induced by nonradiative processes leads to damping of the 2-photon Rabi oscillations.

\subsection{Cavity-shielded 2-photon Rabi oscillation}
We begin by studying the nature of 2-photon non-linear transitions outside vs inside cavities where the $0\rightarrow1$ and $1\rightarrow 2$ transitions are not on resonance with $\omega_{in}$.  In the context of Maxwell-Bloch equations, because the EM field is purely classical and therefore there is no simple conversion between the EM field amplitude and the cavity mode population in the context of a quantum model, we  will not be able to easily measure the populations of the lower and upper polaritons. Therefore, instead, we will calculate the population on molecular layers (the diagonal elements of density matrices), keeping in mind that these molecular states are not eigenstates of the system in the presence of interactions between the cavity EM field and molecular excited states; some oscillations are expected (at early times especially). The frequency of the Rabi oscillations between $\ket{0}$ and $\ket{2}$ is dictated by the rate of 2-photon nonlinear transitions.  To this end, we compare the dynamics for the scenarios without a cavity ((a) and (b)) vs the scenarios with a cavity ((c) and (d)) in Fig \ref{fig: systemsketch}. 
Because (as is well known\cite{siegman1986lasers}) a cavity can trap the EM field, for a fair comparison between outside vs inside cavity, we manually increase the electric field of the external driving for scenarios (a) and (b) by a factor of $3.4$ that was found empirically to be the field enhancement inside the cavity compared with the incident field, i.e. $E_{0}^{a, b} = 3.4\times E_{0}^{c, d}$.

\begin{figure}[ht!]
\includegraphics[width=12cm]{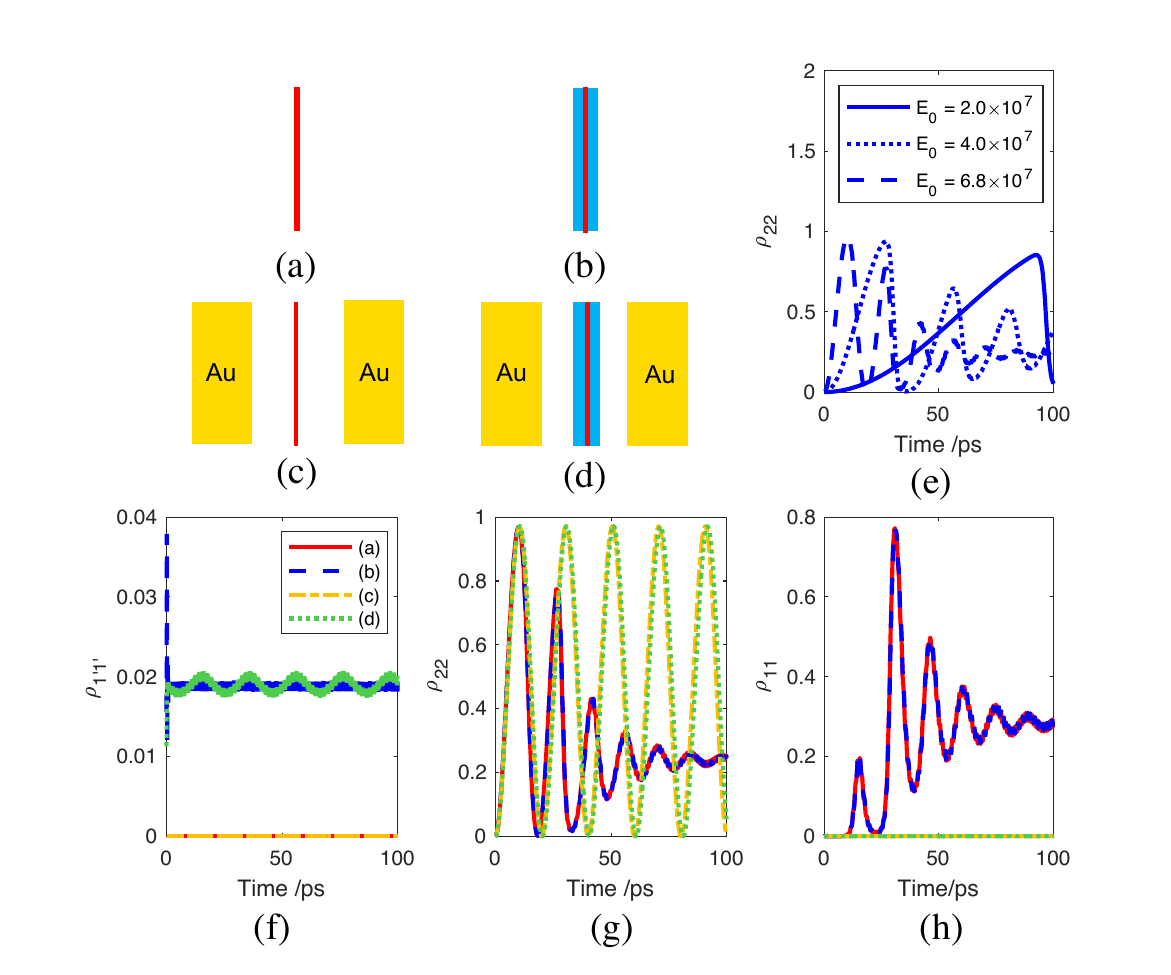}
\caption{(a)-(d) The standard four scenarios under probing with monochromatic light (see Fig. \ref{fig: systemsketch}). (e) Population dynamics in time for state $\ket{2}$ (corresponding to the solute molecular layer) for a series of different intensities with geometry (b). (f)-(h) Population dynamics of the excited states of solvent $\rho_{1'1'}$,  solute $\rho_{22}$ and $\rho_{11}$, respectively. In (f), the red and yellow lines (showing results for scenarios (a) and (c)) overlap with the $X$ axis ($\rho_{1'1'}\equiv0$); for the blue line [scenario (b)], $\rho_{1'1'}$ reaches steady state very fast; for the green line,  corresponding to scenario (d), $\rho_{1'1'}$ oscillates around the steady state value of (b). In (g), for the scenarios outside the cavity [(a) and (b)], $\rho_{22}$ oscillates before reaching a steady state with excitation balanced by emission; for scenarios inside the cavity (c) and (d), $\rho_{22}$ oscillates for at least $100$ ps. Finally, in subfigure (h), in scenarios (a) and (c), we find $\rho_{11}$ reaches a long-time, steady-state limit while, by contrast, for  scenarios (b) and (d), $\rho_{11}$ remains $0$ within $100$ ps of our simulation. These effects can all be rationalized; see the text.}
\label{fig: popvstime4scenarios}
\end{figure}

As a sanity check, in Fig \ref{fig: popvstime4scenarios} (e), we plot the population dynamics of $\rho_{22}$ for scenario (b) with different EM field strengths. As one would expect, stronger EM field strengths result in faster oscillations. 

In general, below, when discussing dynamics in a cavity, we will analyze first the population of the  solvent state  $\ket{1'}$, second the population of state $\ket{2}$, and third the population of state $\ket{1}$ in this order -- because this ordering corresponds to how the states are populated in time under an incoming cw em field.
Let us now analyze our observations about the Rabi oscillations and population dynamics in Fig. \ref{fig: popvstime4scenarios}.

\begin{itemize}
\item Consider first the population dynamics of $\rho_{1'1'}(t)$.  As shown in Fig \ref{fig: popvstime4scenarios} (f),  because there is no solvent molecular layer for (a) and (c), obviously $\rho_{1'1'}\equiv 0$. More interestingly, for scenario (b), the population reaches steady state very fast. Finally,  for scenario (d), there are persistent oscillations about the steady state of (b). In other words, the coherent transitions last longer because of the cavity.

\item Next, consider the population dynamics of $\rho_{22}(t)$. As shown in Fig \ref{fig: popvstime4scenarios} (g), for scenarios (a) and (b), we observe damped Rabi oscillations; for scenarios (c) and (d), we observe persistent Rabi oscillations. The Rabi oscillations for scenario (d) match the oscillations in Fig \ref{fig: popvstime4scenarios}(f); there is clearly energy exchange between the solvent polariton and the solute excitation.  Note also that scenarios (c) and (d) yield identical oscillation patterns. The incoming EM field enters the cavity with the same efficiency for scenario (c) (where $\omega_{in}$ matches the cavity geometry) and scenario (d) (where $\omega_{in}$ matches the lower polariton state of solvent and solute); recall that, according to Fig. \ref{fig: systemsketch}, the cavity lengths are slightly different for scenarios (c) and (d).

\item Lastly, consider the population dynamics $\rho_{11}(t)$. As shown in Fig \ref{fig: popvstime4scenarios} (h), for the scenarios outside the cavity ((a) and (b)), although $E_{01} \neq \hbar\omega_{in}$, $\rho_{11}$ becomes finite because of emission from state $\ket{2}$ to state $\ket{1}$. In contrast, for scenarios inside the cavity ((c) and (d)), $\rho_{11}$ remains $0$.  In other words, we find that inside a cavity, the $\ket{2}\rightarrow\ket{1}$ transition is effectively suppressed by the cavity as $E_{cav}\neq E_{12}$. The steady state population $\rho_{11}(t\rightarrow \infty)$ is proportional to the incoming electric field (in the weak field regime). 
\end{itemize}

\begin{figure}
\includegraphics[width=14cm]{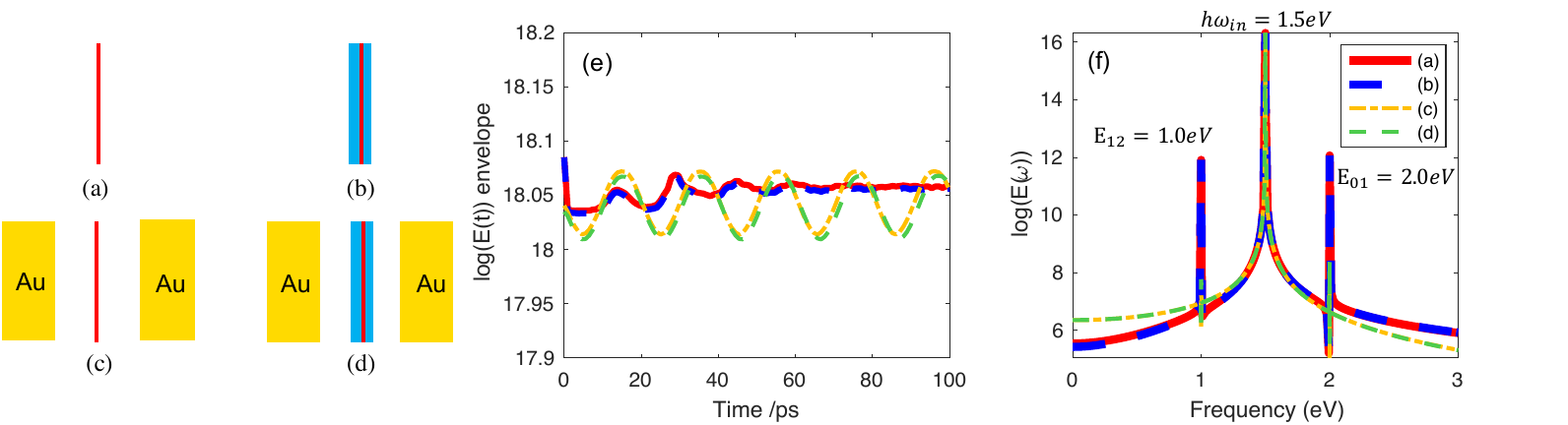}
\caption{(a-d) The standard scenarios considered (see Fig. \ref{fig: systemsketch}).  We plot the electric field envelope acting on the 3-level molecular layer in (e) time and (f) the frequency domain (both in log scale). As shown in Fig. (e), the envelope of the electric field for scenarios inside the cavity (c) and (d) oscillates with an identical frequency as in Fig \ref{fig: popvstime4scenarios} (f) and (h), indicating energy transfer between molecules and the EM field. For scenarios outside the cavity (a) and (b), there is no such oscillation. In the frequency domain, the major peak at $1.5$ eV corresponds to $\hbar\omega_{in}$ and appears in all four scenarios. By contrast, the peaks at $E_{12}= 1.0$ eV and $E_{01} = 2.0$ eV  appear only for scenarios (a) and (b).}
\label{fig: emfieldtimefreq}
\end{figure}

Now, our description above of the process whereby the $2\rightarrow 1$ transition is suppressed by the cavity was inferred from population dynamics. To complement this point of view,  and as mentioned in the introduction, we would like to analyze the corresponding electric field as well using our FDTD simulation.  In Fig. \ref{fig: emfieldtimefreq}, we plot the electric field on the solute molecular layer in both time and frequency domains in logarithm scale. As shown in Fig \ref{fig: emfieldtimefreq} (e),  for scenarios (c) and (d), the continuous Rabi oscillations in Fig \ref{fig: popvstime4scenarios} synchronize with the oscillation of the electric field envelope, which demonstrates that energy is exchanged between the cavity EM field and the solute molecular layer. In contrast, for scenarios (a) and (b), the envelope is practically flat; there is no extra energy exchange after the absorption and emission processes reach a steady state.

Next, we perform Fourier transformations on the signals in (e), and as shown in Fig \ref{fig: emfieldtimefreq} (f), we observe two peaks at $1.0 \text{ and } 2.0$ eV for the scenario with no cavity mirrors. The positions of the peaks correspond to $\ket{2} \rightarrow \ket{1}$ and $\ket{1} \rightarrow \ket{0}$ emissions from the solute molecular layers respectively. 
That being said,  note that Fig. \ref{fig: emfieldtimefreq} is plotted on a log scale and these side peaks are very small. In the end, our conclusion is clear: because these $\ket{2} \rightarrow \ket{1}$ and $\ket{1} \rightarrow \ket{0}$ transitions do not match the cavity geometry, these transitions remain suppressed for scenarios inside the cavity (c) and (d).

\subsection{The role of mirror thickness and quality factor}
In the previous subsection, we compared the population dynamics for the four scenarios (a-d) in Fig. \ref{fig: systemsketch} where we implemented an adjustment to the magnitude of the incoming field ($E_{0}^{a, b} = 3.4\times E_{0}^{c, d}$) which was intended to keep the local electric field roughly the same in and out of the cavity; in other words, for scenarios (a-b) outside the gold cavity, we applied pumping field with amplitude greater than that used to excite the systems for scenarios (c-d) inside the cavity.
In this subsection, we will ignore such a difference and directly probe how the quality factor of the microcavity for scenario (d) can increase/decrease the rate of nonlinear transitions under identical external pumping (($E_{0}^{a, b} \equiv E_{0}^{c, d}$)). 
Note that the quality factor is manifested by the thickness of the gold mirrors ($L = 0-100$nm) in scenario (d) so that,  when $L = 0$, (d) reduces to (b) and there is no cavity; again, we maintain a constant external pumping field strength at all times.
The mirror distance ($d_{2}=340$ nm) is calibrated such that $E_{cav}\approx E_{0'1'}$ so that the lower polariton state will form.
\begin{figure}
\includegraphics[width=15cm]{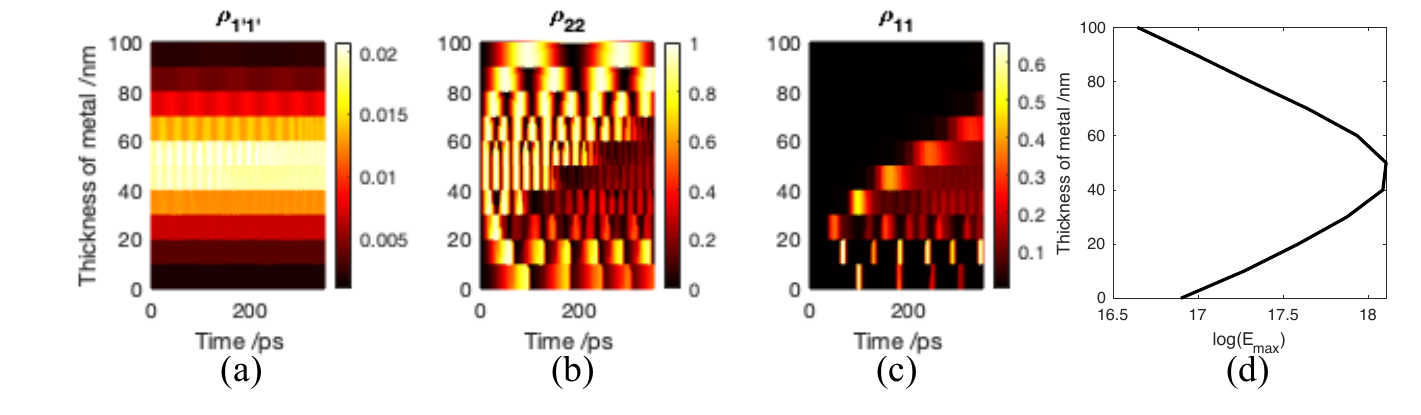}
\caption{Population dynamics on (a) excited state ($\ket{1'}$) of the solvent molecules and (b-c)  excited states $\ket{2}$ and $\ket{1}$ of the solute molecules for scenario (d) in Fig. \ref{fig: systemsketch} with different choices of mirror thickness $L = 0-100$ nm. (d) Maximal circulating electric field amplitude inside the cavity vs mirror thickness (log scale). Note that,  in subfigures (b) and (d), we find an optimal thickness $L\approx50 nm$ for the fastest nonlinear transition corresponding to the maximal circulating electric field amplitude (whose implications are discussed in the text).  The distance between the mirrors is kept constant ($d_{2}\equiv 340$ nm) and in resonance with the solvent molecules ($E_{cav}=E_{0'1'}$) for all choices of mirror thickness. }
\label{fig: popvsthickness}
\end{figure}

In Fig \ref{fig: popvsthickness}, we scan over different thicknesses for the mirror ($L$) and compare the resulting population dynamics ($\rho_{1'1'}(t)$, $\rho_{11}(t)$ and $\rho_{22}(t)$). 
\begin{itemize}
    \item In Fig \ref{fig: popvsthickness}(a), we plot the time dynamics for the excited state population of the 2-level solvent molecular layers $\rho_{1'1'}(t)$ for a varying mirror thickness ($L = 0-100$ nm, $Y$ axis). We find that $\rho_{1'1'}(t)$ reaches a steady state because of the balance between the external pumping and the decay induced by cavity leakage/mirror absorption. Moreover, as the mirror thickness increases, the steady state value reaches a maximum near a $40-60$ nm mirror thickness and becomes smaller for thicker mirrors. As shown in Fig \ref{fig: popvsthickness} (d), this maximum corresponds to the greatest circulating electric field amplitude inside the cavity at steady state.
    \item In Fig \ref{fig: popvsthickness}(b), we plot the population of the second excited state of the 3-level solute molecular layers $\rho_{22}(t)$. We find that $\rho_{22}(t)$ undergoes Rabi oscillations for all choices of the mirror thicknesses (note that similar oscillations were also found in Fig \ref{fig: popvstime4scenarios}(g)). In agreement with subfigure (a), we observe that the frequency of the Rabi oscillation shows a non-monotonic behavior. Starting from a cavity with a very low quality factor ($L=10 $nm), the Rabi frequency increases when the mirror thickness increases. \footnote{Note that, the cavity mode energy is never strictly $1.5$ eV as we vary mirror thicknesses. Here we choose to keep a constant distance between two mirrors for consistency.} However, when the quality factor becomes very large ($L=100 $nm), the external EM field cannot enter the cavity efficiently such that the effective amplitude of the pumping field inside the cavity near the molecular layers is reduced (and the Rabi frequency decreases). Note that the Rabi oscillations do not last forever. For example, for mirror thickness $L = 50 $nm, the Rabi oscillations stop after $200$ ps. In other words, as the mirrors thickness $L$ increases, the Rabi oscillations last longer until $\rho_{22}$ decays to $\rho_{11}$ as induced by the cavity leakage (discussed in the following).
    \item In Fig \ref{fig: popvsthickness}(c), we plot the population of the first excited state of the solute molecular layers, $\rho_{11}(t)$.
    For low-quality factors ($L < 40$ nm), even though $E_{01}\neq E_{in}$, $\rho_{11}$ is not zero for the same reason as in Fig. \ref{fig: popvstime4scenarios} (h) (corresponding to scenario (d)), we also observe that the population $\rho_{11}$ becomes non-zero only after at least one cycle of the Rabi oscillation of $\rho_{22}$ in (b). This observation suggests that the population $\rho_{11}$ arises from the transition from $\ket{2}$ to $\ket{1}$. 
    As the thickness of the mirrors becomes larger ($40< L < 60$ nm), it takes more Rabi oscillation cycles of $\rho_{22}$ to make such a transition to populate $\rho_{11}$ and, after the transition, the maximal population of $\rho_{22}$ drop while maintaining the Rabi oscillation. When $L > 60$ nm, we find that $\rho_{11}(t \leq 100 \text{ps}) = 0$.
\end{itemize}

The bottom line is that these plots show that increasing the quality factors (i.e. making the mirror thicker) can lead to non-monotonic behavior: there is a trade-off insofar as collective response being enhanced when a photon bounces back and forth in a cavity versus collective behavior being reduced when a  photon cannot enter the cavity to begin with.

\subsection{Dynamics with phenomenological pure dephasing}
\begin{figure}
\includegraphics[width=15cm]{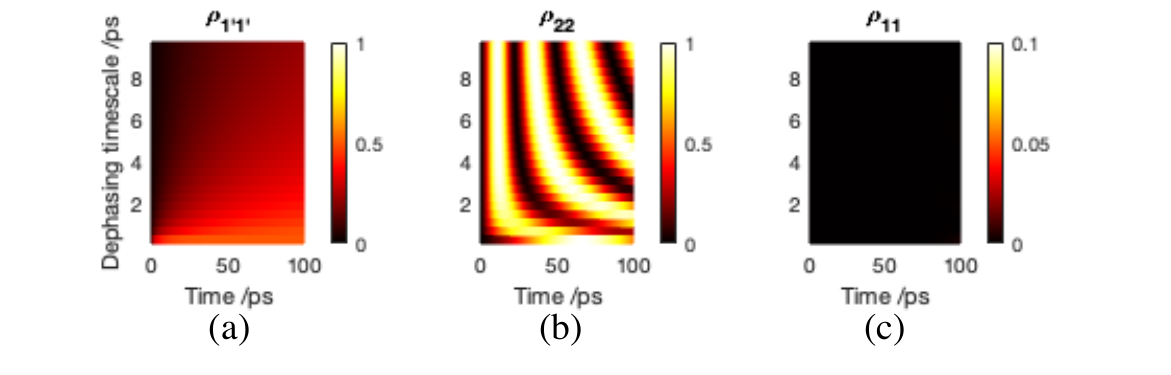}
\caption{Population dynamics (a) $\rho_{1'1'}(t)$, (b) $\rho_{11}(t)$ and (c) $\rho_{22}(t)$ when we dephase  the solvent molecular layers for scenario (d) ($\tau_{d} = 1-10$ ps). For slow dephasing, the Rabi splitting remains unperturbed. When the dephasing rate is increased (corresponding to a smaller dephasing timescale in the lower part of the figures), the excited state population $\rho_{1'1'}$ of the solvent molecules is trapped, leading to ground state depletion and Rabi splitting contraction. Hence, given the identical incoming pumping frequency $\hbar\omega_{in} = 1.5$ eV, the external pumping cannot enter the cavity efficiently and $\rho_{11}$ is not populated within the first 100 ps.}
\label{fig: popvstimedecoherence1}
\end{figure}
\begin{figure}
\includegraphics[width=15cm]{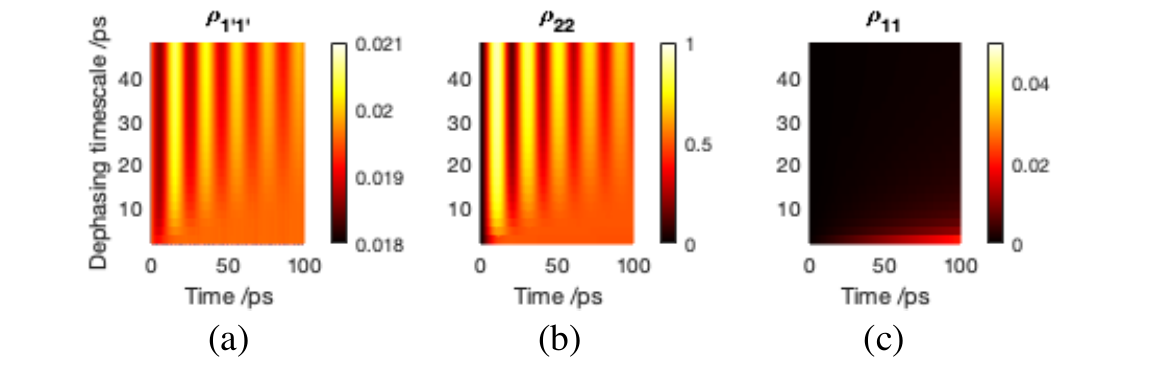}
\caption{Population dynamics (a) $\rho_{1'1'}(t)$, (b) $\rho_{11}(t)$ and (c) $\rho_{22}(t)$ when we dephase the 3-level solute molecular layer for scenario (d) ($\tau_{d} = 1-10$ ps). Difference arises versus  Fig \ref{fig: popvstimedecoherence1} because now the dephasing is on the solute molecules.  We find that the Rabi oscillation pattern in (b) is damped  in the strong dephasing limit and the system reaches steady state at $\rho_{22} = 0.5$. The time dynamics of the solvent molecules remains mostly the same as for scenario (d) in Fig \ref{fig: popvstime4scenarios} (f), except for the fact that the persistent small oscillation is damped in the strong dephasing limit. As in Fig. \ref{fig: popvstimedecoherence1}(c), we do not see any substantial population in $\rho_{11}$.}
\label{fig: popvstimedephasing2}
\end{figure}

In practice, the 2-photon Rabi oscillation pattern does not last forever as shown in \ref{fig: popvsthickness}.
The Rabi oscillations can be damped by other non-radiative processes and dephasing. In this subsection, we investigate how pure dephasing as induced by the environment changes the time dynamics of scenario (d). The pure dephasing is implemented by manually damping the coherence $\rho_{01}$, $\rho_{02}$, $\rho_{0'1'}$ and $\rho_{12}$ by a factor of $\exp(-dt/\tau_{d})$ for each timestep $dt$. For a typical experiment, the pure dephasing timescale ($\tau_{d}$) is between $1$ ps to $100$ ps. To best isolate the effects of pure dephasing, we perform two parallel sets of calculations with dephasing either (i) exclusively on the solvent molecular layers, as shown in Fig \ref{fig: popvstimedecoherence1}, or (ii) exclusively on the solute molecular layer, as shown in Fig \ref{fig: popvstimedephasing2}. In practice, pure dephasing should apply to both molecular layers with similar rates, 
but the following gedanken experiments will make clear different effects.

First, for pure dephasing on the solvent molecular layers, the direct effect is that the radiative decay from $\ket{1'}$ to the ground state $\ket{0'}$ is suppressed, which leads to ground state depletion and Rabi splitting contraction. Hence, the resonant frequency of lower polariton is no longer the same as the incoming driving ($E_{LP} > E_{in}$). As a result, the maximal electric field amplitude decreases inside the cavity and the Rabi oscillations of the solute molecular layer between $\ket{0}$ and $\ket{2}$ are slowed down, as shown in Fig \ref{fig: popvstimedecoherence1} (b).

Second, in Fig \ref{fig: popvstimedephasing2} we investigate dynamics with pure dephasing on the solute layer. As shown in (a), as the dephasing rate increases, the small oscillations of $\rho_{1'1'}$ (see Fig \ref{fig: popvstime4scenarios}(f)) are damped, yet the population remains around the same steady state value.  Similarly, as far as $\rho_{22}$ is concerned, the Rabi oscillation pattern is damped (but not slowed down). Dephasing the solute is not as interesting as dephasing the solvent.

\section{Conclusion and Outlook\label{sec: four}}
In conclusion, we have performed finite difference time domain (FDTD) calculations  within a mean field approximation to simulate 2-photon nonlinear transitions between a continuous wave light and 3-level solute molecular layers. The population dynamics of the molecular layers are obtained outside and inside a one-dimensional cavity made of two gold mirrors. We find that the presence of a cavity changes the dynamics in two aspects: (i) the cavity amplifies the EM field such that the transitions occur faster when the quality factor is moderately large; and (ii) the cavity inhibits transitions that are not on resonance, which effectively protects the coherent Rabi oscillation dynamics from relaxation. However, if we further increase the quality factor of the cavity, the external electromagnetic field cannot enter the cavity efficiently, leading to slower Rabi oscillations. We also investigate the role of dephasing on the solute and the solvent layers and, in short, both types demolish the two-photon transition.

Looking forward, one can envision several improvements to the current calculations. First, 
it is well-known that under a mean field assumption (where the EM field is classical and modulates all interactions between molecules), some elements of static intermolecular coupling are not included.
A proper configuration interaction  Hamiltonian that incorporates intermolecular coupling may well reveal new physics -- but this approach would certainly be more expensive. Second, the current calculations lack any (even phenomenological) nonradiative decay between excited states and ground states. 
In the future, one might include such an effect by introducing a  Lindblad term into the equation of motion; such a term would certainly damp the Rabi oscillations.
Third and finally, we note that the calculations presented here are only one-dimensional: the cavity has multiple modes but only along one polarization. Practically speaking, our setup corresponds to concentric cavities, i.e. there is effectively a non-zero curvature on the mirrors.
In the future, one would like to perform two/three-dimensional FDTD calculations to include more polarization directions.

Beyond improving the present calculations, ideally the next step of this research is to go beyond two-level systems and model realistic systems with realistic interactions between EM fields molecular layers inside a multi-mode cavity. For example, in the future, one would like to generalize the present set of simulations to study the behavior of matter in a cavity under very strong photonic excitation where, for example, BEC phases appear\cite{shelykh2006polarization, deng2007spatial, byrnes2014exciton}. The intersection of electronic dynamics and strong coupling to light fields form a very rich area for future exploration.

\section*{Acknowledgements}
This work has been supported by the U.S. Department of
Energy, Office of Science, Office of Basic Energy Sciences,
under Award No. DE-SC0019397 (J.E.S.); the U.S. National
Science Foundation under Grant No. CHE1953701 (A.N.). 
MS acknowledges support by the Air Force Office of Scientific Research under Grant No. FA9550-22-1-0175.

\appendix
\section{Parameters for FDTD simulation}
In this appendix, we list all parameters in our simulation.
\begin{table}[ht!]
  \begin{threeparttable}
   \caption[]{Parameters for Maxwell-Bloch simulation.}
   \centering
   
   \begin{tabular}{P{8cm}c}
     \midrule 
     Name & Value\tnote{$\dagger$}
    \\\hline
    Grid resolution ($dx$) & $0.1$ nm
    \\
    Time step (dt) & $dx/2/c_{0}$
    \\
    Thickness of solute & $1$ nm
    \\
    Thickness of solvent & $4.5$ nm $\times 2$
    \\
    Thickness of gold & $50$ nm $\times 2$
    \\
    Distance between gold mirrors in Fig \ref{fig: systemsketch}(c) & $354$ nm
    \\
    Distance between gold mirrors in Fig \ref{fig: systemsketch}(d) & $340$ nm
    \\
    Drude model for gold & $\omega_{d} = 7.039$eV
    \\
    & $\Gamma_{d} = 0.1809$ eV
    \\
    Transition dipole moment & $\mu_{x}^{01}=\mu_{x}^{12}=\mu_{x}^{0'1'} = 10$ Debye
    \\
    
    \midrule
     \end{tabular}
  \end{threeparttable}
\end{table}

\bibliography{apssamp}

\end{document}